\documentclass[fleqn,usenatbib]{mnras}

\usepackage[T1]{fontenc}
\usepackage{ae,aecompl}

\usepackage{graphicx}

\usepackage{nohyperref}

\title[Machine Learning for Transit Detection]{Machine-learning Approaches to Exoplanet Transit Detection and Candidate Validation in Wide-field Ground-based Surveys}

\author[N. Schanche et al]{
N. Schanche,$^{1}$ \thanks{E-mail: ns81@st-andrews.ac.uk}
A. Collier Cameron,$^{1}$ 
G. H{\'e}brard,$^{2}$
L. Nielsen,$^{3}$ \newauthor
A. H.M.J. Triaud,$^{4}$ 
J.M. Almenara, $^{5}$
K.A. Alsubai, $^{6}$
D.R. Anderson,$^{7}$ \newauthor
D.J. Armstrong,$^{8,9}$
S.C.C. Barros,$^{10}$ 
F. Bouchy,$^{3}$
P. Boumis,$^{11}$ 
D.J.A. Brown,$^{8,9}$ \newauthor
F. Faedi,$^{9,12}$ 
K. Hay,$^{1}$ 
L. Hebb,$^{13}$ 
F. Kiefer,$^{2}$
L. Mancini,$^{14,15,16,17}$ 
P.F.L. Maxted,$^{7}$\newauthor
E. Palle,$^{18,19}$ 
D.L. Pollacco,$^{8,9}$
D. Queloz,$^{3,20}$ 
B. Smalley,$^{7}$ 
S. Udry,$^{12}$
R. West,$^{8,9}$ \newauthor
P.J. Wheatley$^{8,9}$
\\ 
\\
$^{1}$Centre for Exoplanet Science, SUPA, School of Physics and Astronomy, University of St Andrews, St Andrews KY16 9SS, UK\\
$^{2}$Institut d'astrophysique de Paris, UMR7095 CNRS, Universit\'e Pierre \& Marie Curie, 98bis boulevard Arago, 75014 Paris, France\\
$^{3}$Observatoire de Genève, Université de Genève, 51 Chemin des Maillettes, CH-1290 Sauverny, Switzerland\\
$^{4}$School of Physics \& Astronomy, University of Birmingham, Edgbaston, Birmingham B15 2TT, UK \\
$^{5}$Universit\'e Grenoble Alpes, CNRS, IPAG, 38000 Grenoble, France\\
$^{6}$Qatar Environment and Energy Research Institute (QEERI), Hamad Bin Khalifa University (HBKU), Qatar Foundation, Doha, Qatar \\
$^{7}$Astrophysics Group, Keele University, Staffordshire, ST5 5BG, UK\\
$^{8}$Centre for Exoplanets and Habitability, University of Warwick, Gibbet Hill Road, Coventry, CV4 7AL, UK \\
$^{9}$Department of Physics, University of Warwick, Coventry CV4 7AL, UK\\
$^{10}$Instituto de Astrof\'isica e Ci\^{e}ncias do Espa\c co, Universidade do Porto, CAUP, Rua das Estrelas, 4150-762 Porto, Portugal\\
$^{11}$Institute for Astronomy, Astrophysics, Space Applications and Remote Sensing, National Observatory of Athens, 15236 Penteli, Greece\\
$^{12}$INAF - Osservatorio Astrofisico di Catania, Via S. Sofia 78, I-95123 Catania, Italy\\
$^{13}$Hobart and William Smith Colleges, Department of Physics, Geneva, NY 14456, USA\\
$^{14}$Department of Physics, University of Rome Tor Vergata, Via della Ricerca Scientifica 1, 00133 Roma, Italy\\
$^{15}$Max Planck Institue for Astronomy, K\"{o}nigstuhl 17, 69117 Heidelberg, Germany\\
$^{16}$INAF - Astrophysical Observatory of Turin, Via Osservatorio 20, 10025 Pino Torinese, Italy\\
$^{17}$International Institute for Advanced Scientific Studies (IIASS),Via G. Pellegrino 19, I-84019 -- Vietri sul Mare (SA), Italy \\
$^{18}$Instituto de Astrosf\'isica de Canarias (IAC), 38205 La Laguna, Tenerife, Spain\\
$^{19}$Departamento de Astrof\'isica, Universidad de La Laguna (ULL), 38206 La Laguna, Tenerife, Spain\\
$^{20}$Cavendish Laboratory, JJ Thompson Avenue, CB3 0HE, Cambridge, UK
}

\date{Accepted 2018 November 13. Received 2018 October 22; in original form 2018 August 15}

\pubyear{2018}

\begin{document}
\label{firstpage}
\pagerange{\pageref{firstpage}--\pageref{lastpage}}
\maketitle

\begin{abstract}

Since the start of the Wide Angle Search for Planets (WASP) program, more than 160 transiting exoplanets have been discovered in the WASP data. In the past, possible transit-like events identified by the WASP pipeline have been vetted by human inspection to eliminate false alarms and obvious false positives. The goal of the present paper is to assess the effectiveness of machine learning as a fast, automated, and reliable means of performing the same functions on ground-based wide-field transit-survey data without human intervention. To this end, we have created training and test datasets made up of stellar light curves showing a variety of signal types including planetary transits, eclipsing binaries, variable stars, and non-periodic signals. We use a combination of machine learning methods including Random Forest Classifiers (RFCs) and Convolutional Neural Networks (CNNs) to distinguish between the different types of signals. The final algorithms correctly identify planets in the test data $\sim$90\% of the time, although each method on its own has a significant fraction of false positives. We find that in practice, a combination of different methods offers the best approach to identifying the most promising exoplanet transit candidates in data from WASP, and by extension similar transit surveys.

\end{abstract}


\section{Introduction}
Exoplanet transit surveys such as the Convection Rotation and Planetary Transits \citep[CoRoT;][]{Auvergne2009}, Hungarian-made Automated Telescope Network \citep[HATnet;][]{Hartman2004}, HATSouth \citep{Bakos2013}, the Qatar Exoplanet Survey \citep[QES;][]{Alsubai2013}, the Wide-angle Search for Planets \citep[WASP;][]{Pollacco2006}, the Kilodegree Extremely Little Telescope \citep[KELT;][]{Pepper2007}, and \textit{Kepler} \citep{Borucki2010} have been extremely prolific in detecting exoplanets, with over 2,900 confirmed transit detections as of August 9, 2018\footnote{\url{https://exoplanetarchive.ipac.caltech.edu/index.html}}. 

The majority of these surveys employ a system where catalogue-driven photometric extraction is performed on calibrated CCD images to obtain an array of light curves. Following decorrelation of common patterns of systematic error (eg \cite{Tamuz2005}), an algorithm such as the Box-Least Squares method \citep{Kovacs2002} is applied to all of the lightcurves. Objects that have signals above a certain threshold are then identified as potential planet candidates. Before a target can be flagged for follow-up observations, the phase-folded light curve is generally inspected by eye to verify that a genuine transit is present. As these surveys contain thousands of objects, the manual component quickly becomes a bottleneck that can slow down the identification of targets. Additionally, even with training it is difficult to establish consistency in the validation process across different observers. It is therefore desirable to design a system that can consistently identify large numbers of targets more quickly and accurately than the current method.

Several different techniques have been used to try to automate the process of planet detection. A common method is to apply thresholds to a variety of different data properties such as signal-to-noise ratio, stellar magnitude, number of observed transits, or measures of confidence of the signal, with items exceeding the given threshold being flagged for additional study (For WASP-specific examples, see \cite{Gaidos2014} and \cite{Christian2006}). Applying these criteria can be a fast and efficient way to find specific types of planets quickly, but they are not ideal for finding subtle signals that cover a wide range of system architectures. 

Machine learning has quickly been adopted as an effective and fast tool for many different learning tasks, from sound recognition to medicine (See, e.g., \cite{Lecun2015} for a review). 
Recently, several groups have begun to use machine learning for the task of finding patterns in astronomical data, from identifying red giant stars in asteroseismic data \citep{Hon2017} to using photometric data to identify quasars \citep{Carrasco2015}, pulsars \citep{Zhu2014}, variable stars \citep[]{Pashchenko2018, Masci2014, Naul2017, Dubath2011, Rimoldini2012}, and supernovae \citep{duBuisson2015}.  For exoplanet detection in particular, Random Forest Classifiers \citep{McCauliff2015,Mislis2016}, Artificial Neural Networks \citep{Kipping2017}, Convolutional Neural Networks \citep{Shallue2018}, and Self-Organizing Maps \citep{Armstrong2017} have been used on \textit{Kepler} archival data. Convolutional Neural Networks were trained on simulated \textit{Kepler} data by \cite{Pearson2018}.

While \textit{Kepler} provides an excellent data source for machine learning (regular observations, no atmospheric scatter, excellent precision, large sample size), similar techniques can also be applied to ground-based surveys, and in fact machine learning techniques have recently been incorporated by the MEarth project \citep{Dittmann2017} and NGTS \citep{Armstrong2018}. We extend this work by applying several machine learning methods to the WASP database. In section \ref{Observations} we briefly describe the current process for WASP candidate evaluation. Then in section \ref{Methods} we discuss the methods developed, focusing on Random Forest Classification (RFC) and Convolutional Neural Networks (CNN), and describe how these methods are applied to data from the WASP telescopes. In section \ref{Analysis} we discuss the efficacy of the machine-learning approach, with an emphasis on the false-positive candidate identification rate. Finally, in sections \ref{Discussion} and \ref{Conclusions}, we discuss practical applications of the machine classifications in the future follow-up of planetary candidates.

\section{Observations} \label{Observations}

For this work, we focus entirely on WASP data, although similar techniques would be applicable to any ground-based wide-field transit survey. The WASP project \citep{Pollacco2006} consists of two robotic instruments, one in La Palma and the other in South Africa. The project was designed using existing commercial components to reduce costs, so each location is made up of 8 commercial cameras mounted together and using science-grade CCDs for imaging.  

The WASP field is comprised of tiles on the sky corresponding approximately to the 7.8 degree square field of view of a single WASP camera. The WASP database includes data on all objects in the tiles secured over all observing seasons in which the field was observed. Decorrelation of common patterns of systematic error has been carried out using a combination of the Trend Filtering Algorithm \citep[TFA;][]{Kovacs2005} and the SysRem algorithm \citep{Tamuz2005}. Initial data reduction was carried out season-by-season and camera-by-camera. More recently, re-reduction with the ORCA\_TAMTFA (ORion transit search Combining All data on a given target with TAMuz and TFA decorrelation) method has yielded high-quality final light curves for all WASP targets. Each tile typically contains between 10000 and 30000 observations.

In total, there are 716 ORCA\_TAMTFA fields, with each field containing up to 1000 tentative candidates identified as showing Box Least Squares (BLS) signals above set thresholds in anti-transit ratio and signal-to-red-noise ratio (see \cite{CollierCameron2006} for details). 

At this point in the process, WASP targets are selected for follow-up by a team of human observers. The observers can either look at lightcurves by field or can apply filters, cutting on thresholds for selected candidate properties. Finding these cuts has been done through trial and error, and can vary between observers. The targets of interest are prioritized for radial-velocity or photometric follow-up observations. Only after successful secondary observations can an object be identified as a planet in the database. However, both the vetters and the follow-up observers can flag the star as something else, such as an eclipsing binary, stellar blend, or variable star before or after follow-up observations confirm the source, which makes misclassifications of these object types possible. The final categorization (planet, eclipsing binary, variable, blend, etc) recorded by human vetters or observers is known as the disposition. 

While effort was taken early in the WASP project to standardize individual classifications through training sessions and cross validation, the current method of identifying planetary candidates remains partially dependent on individual opinions. It would be better to establish a system that can systematically go through all of the data and identify targets, ideally ranked by their likelihood to be genuine planets, derived from the knowledge gained from the entire history of past dispositions. The past decade of classifications and observations has generated a dataset containing both descriptions of the target and their classification, creating an excellent starting point for supervised machine learning.


\section{Methods} \label{Methods}

Several classification algorithms were explored with the goal of reliably identifying planet candidates. We seek to more efficiently use the telescope time for follow-up observations by reducing the number of false positive detections as far as is reasonable without compromising sensitivity to rare classes of planet such as short-period and inflated hot Jupiters. Other goals, such as finding specific subsets of planets with higher precision, could be carried out by retraining the algorithms for that given purpose. In this section we will discuss the different machine learning methods utilized and the datasets created to train the algorithms. 

\subsection{Initial Data Exploration} \label{IDE;}

Periodic signals in the lightcurve can come from astrophysical sources other than transiting planets \citep{Brown2003}. We refer to a periodic dip caused by something other than a planet as an astrophysical false positive. It is essential for any machine learning application to distinguish between planetary signals and false positives. Fortunately many types of astrophysical configurations have been identified in the WASP archive. The training dataset used for the RFC is composed of a table of data containing all of the planets (P), eclipsing binaries, both on their own and blended with other nearby objects (EB/Blend), variable stars (V), and lightcurves containing no planetary transit after human inspection (X). Blends are especially common in WASP data because it has a 3.5-pixel (48 arcsec) aperture, leading to the blending of signals from several stars. Is is notable that the WASP archive labels a planet as 'P' even if discovered by a different survey. While not all of these planets are detectable by eye in the WASP data, planets discovered by other instruments are included in the training sample with the aim of extending the parameter space to which the classifiers are sensitive. Low-mass eclipsing binaries (EBLMs) were excluded from the training and testing datasets, as their signals look photometrically similar to that of a transiting planet. However, we do test the final algorithms performance on these objects in section \ref{Analysis}. 

\begin{figure}
\centering
\includegraphics[width=\columnwidth]{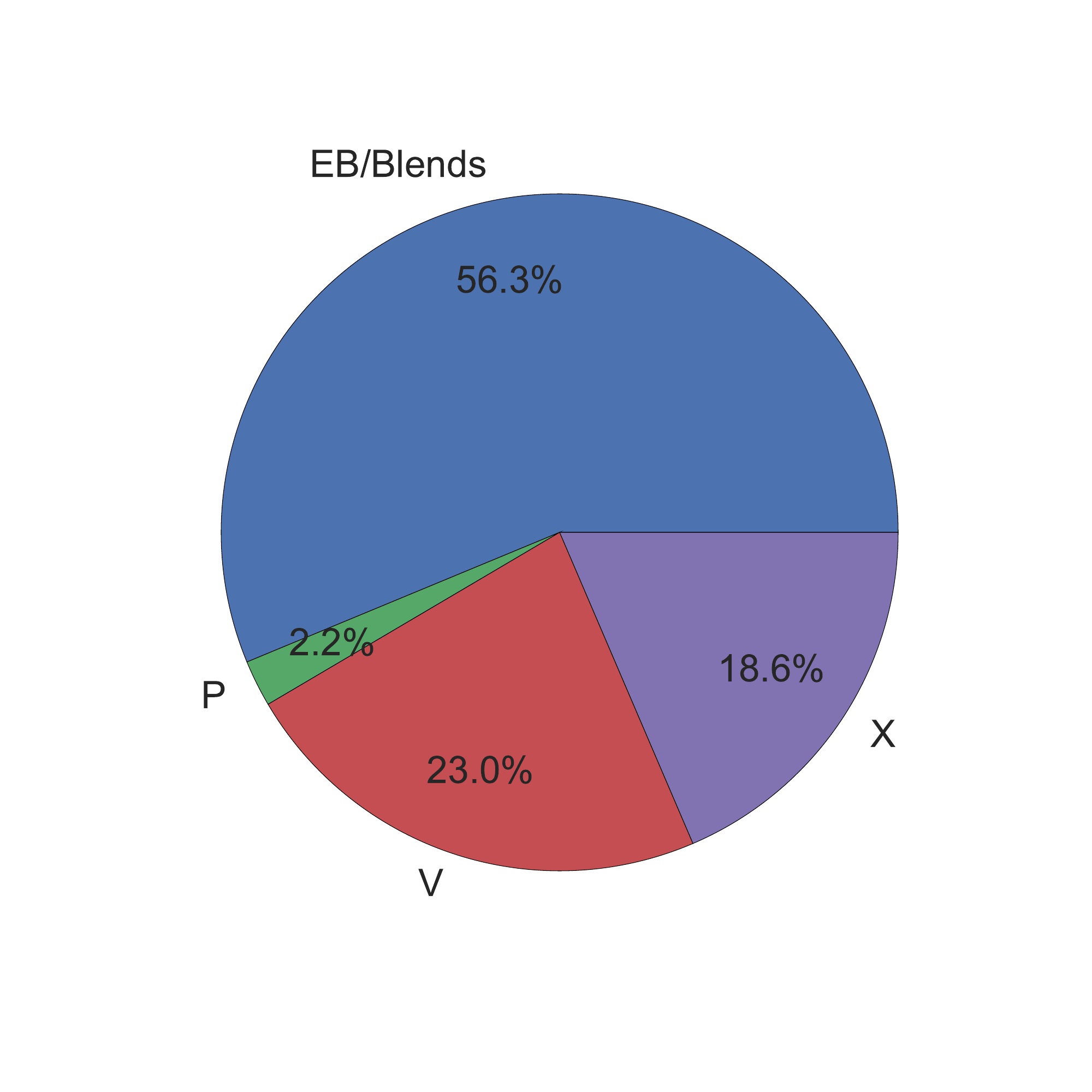}
\caption{Pie plot showing the relative number of examples for planets (P), eclipsing binaries and blends (EB/Blends), variable stars (V) and lightcurves otherwise rejected as planet hosts (X). This represents an unbalanced dataset, with the smallest class being the class of special interest, planets. In real life, there are more non-detections (X) than the other classes. However, the training data is composed of lightcurves that have been manually identified and flagged. Many non-detections are simply ignored by observers rather than labeled X, leading this class to be smaller than expected. The mismatched sample size is important in machine learning, as considerations must be taken to bring representation to the minority classes.} 
\label{fig:RF_pie}
\end{figure}

The final size of each of these classes is shown in Fig. \ref{fig:RF_pie}. All lightcurves used in this study have been classified by members of the WASP team as of August 6, 2018 and have a V-magnitude of less than 12.5. We used only the amalgamated light curves combined across all cameras and observing seasons for which data was present that were then de-trended with a combination of SysRem \citep{Tamuz2005} and the Trend-Filtering Algorithm \citep[TFA;][]{Kovacs2005} and searched with a hybrid Box Least-Squares (BLS) method \citep{Kovacs2002}.

An initial transit width, depth, period, epoch of mid-transit, and radius are estimated from the BLS. Stellar features such as the mass, radius, and effective temperature are found by the method described by \cite{CollierCameron2007}, in which the effective temperature is estimated from a linear fit to the 2MASS $J-H$ color index. The main-sequence radius is derived from $T_{\rm eff}$ using the polynomial relation of \cite{Gray1992}, and the mass follows from a power-law approximation to the main-sequence mass-radius relation, $M_*\propto R_*^{1/0.8}$. A more rigorous fit to the transit profile yields the impact parameter and the ratio of the stellar radius to the orbital separation, and hence an estimate of the stellar density. Markov-chain Monte Carlo (MCMC) runs are performed to sample the posterior probability distributions of the stellar and planetary radii and orbital inclination. The MCMC scheme uses optional Bayesian priors to impose a main-sequence mass and radius appropriate to the stellar effective temperature. Note that the results and predictions would change if the precise radius were used instead, particularly if the star has evolved off of the main sequence.

In addition to the provided information, we add several new features to capture more abstract or relational information, such as the ratio of transit depth to width and the skewness of the distribution of the magnitudes found within the transit event. The latter is a possible discriminator between `U' shaped central transits of a small planet across a much larger star, and shallow `V' shaped eclipses of grazing stellar binaries. The new high precision distance calculations released by \textit{Gaia} \citep[]{Gaia2016, Gaia2018} are used to measure the deviation of the estimated main sequence radius calculated as above and the measured radius. In total, 34 features are included in the dataset. A full list of features and their definitions can be found in Table \ref{table:feature_table}.

\begin{table*}
\caption{Features used by the classifiers. Starred features are those added to the dataset, while the rest were taken directly from the Hunter query. The efficacy of many of these measures for false-positive identification is discussed in detail by \protect\cite{CollierCameron2006}}
\centering 
\begin{tabular}{l l}
\hline
\hline                      
Feature Name & Description \\[0.5ex]
\hline  

clump\_idx & Measure of the number of other objects in the same field with similar period and epoch \\
dchi\_P* & The $\Delta\chi^2$ value at the best-fit period from the BLS method. \\
dchi\_P\_vs\_med* & The ratio of $\Delta\chi^2$ at the best-fit period to median value. \\
dchisq\_mr & Measure of the change in the $\chi^2$ when MCMC algorithm imposes a a main-sequence (MS) prior for mass and radius.  \\ 
delta\_Gaia* & stellar radius from MCMC - Gaia dr2 radius divided by Gaia dr2 radius \\
delta\_m* & The difference between the mass calculated by J-H and the MCMC mass. \\
delta\_r* & The difference between the radius calculated by J-H and the MCMC mass. \\
depth & The depth of the predicted transit from Hunter. \\
depth\_to\_width* & Ratio of the Hunter depth and width measures. \\
epoch & Epoch of the predicted transit from Hunter (HJD-2450000.0) \\
impact\_par & impact parameter estimated from MCMC algorithm. \\
jmag-hmag & Color index, J magnitude - H magnitude. \\
kurtosis* & Measure of the shape of the dip for in-transit data points.  \\
mstar\_jh & Mass of the star, from the J-H radius*(1/0.8). \\
mstar\_mcmc & Stellar mass determined from MCMC analysis. \\
near\_int* & Measure of nearness to integer day periods, abs(mod(P+0.5,1.0)-0.5). \\
npts\_good & Number of good points in the given lightcurve. \\
npts\_intrans & Number of datapoints that occur inside the transit. \\
ntrans & Number of observed transits. \\
period & Detected period by Hunter? in seconds. \\
rm\_ratio* & Ratio of the MCMC derived stellar radius to mass. \\
rplanet\_mcmc & Radius of the planet, from MCMC analysis. \\
rpmj & Reduced proper motion in the J-band (RPMJ=Jmag+5*log10(mu)). \\
rpmj\_diff & Distance from DWs curve separating giants from dwarfs.  \\
rstar\_jh & Radius of the star derived from the J-H color measure. \\
rstar\_mcmc & Radius of the star determined from MCMC analysis \\
sde & Signal Detection Efficiency from the BLS. \\
skewness* & Measure of the asymmetry of the flux distribution of data points in transit. \\
sn\_ellipse & Signal to noise of the ellipsoidal variation. \\
sn\_red & Signal to red noise.  \\
teff\_jh & Stellar effective temperature, from J-H color measure. \\
trans\_ratio & Measure of the quality of data points (data points in transit/total good points)/transit width.  \\
vmag & Cataloged V magnitude. \\
width & Width of the determined transit in hours.  \\ [1ex] 
\hline
\end{tabular}
\label{table:feature_table}
\end{table*}

Before training, the full dataset containing the star name, descriptive features, and disposition is split randomly into a training dataset and a test dataset. In total there are 4,697 training cases and 2,314 testing samples. Prior to running the classifiers, all of the features of the training dataset are median centered and scaled in order to reduce the dynamic range of individual features and to improve performance of the classifier. The scaling parameters are retained so that they can be applied to subsequent datasets to which the classification is applied, including the testing dataset. 

The training dataset is used as an input to a variety of classifiers, namely a Support Vector Classifier \citep[SVC;][]{Cortes1995}, Linear Support Vector Classifier (LinearSVC), Logistic regression \citep[For the implementation is sci-kit learn, see][]{Yu2011}, K-nearest neighbors \citep[KNN;][]{Cover1967}, and Random Forest Classifier \citep[RFC;][]{Breiman2001}. All of the classifiers are run using the relevant functions in Python's \textsc{Scikit-learn} package \citep{Pedregosa2011}. The classification algorithms have default tuning parameters, but they are not always the best choice for the given dataset. We therefore vary the parameters in order to find an optimal combination. While it is impractical to test every possible combination of parameters, we ran a grid of tuning parameter combinations specific to each classification method to find the optimal settings for the final algorithm. The results using the best performing parameters for each method are reported in Table \ref{table:classification_algorithms}. A more detailed description of what each parameter does can be found in the documentation for \textsc{Scikit-learn} \footnote{\url{http://scikit-learn.org/stable/}}.

Several of the classifiers, and particularly the K-nearest neighbor classifier, show poor performance using the training data because the class of interest, the planets, is underrepresented in the dataset. To try to compensate for this, we also try adding additional datapoints using the Synthetic Minority Over-sampling Technique \citep[SMOTE;][]{Chawla2002}. This technique creates synthetic datapoints for the minority classes that lie between existing datapoints with some added random variation. The synthetic data is added only to the training data, and the test dataset remains the same as before. The addition of SMOTE data generally increased the number of planets retrieved from the data, but also increased the number of non-planets given a planet classification. The exception is the SVC, which shows a sharp decrease in false positives while the true positives and true negatives remain the same. 

\begin{table*}

\centering 
\begin{tabular}{l c c c c c c c c}
\hline
\hline                      

Classifier & TP & FP & FN & Precision & Recall & F1 & Accuracy & Tuned Parameters \\
\hline
LinearSVC & 35 & 83 & 13 & 30 & 73 & 43 & 80 & C=60, tol=0.0005\\
SVC & 37 & 46 & 11 & 45 & 77 & 57 & 82 & kernel='rbf', C=12, gamma=0.03\\
LogisticRegression & 34 & 76 & 14 & 31 & 71 & 43 & 81 & C=90, tol=0.005\\
KNN & 5 & 4 & 43 & 56 & 10 & 17 & 81 & n\_neighbors=15, weights='distance'\\
RandomForest & 44 & 125 & 4 & 26 & 92 & 41 &  79 & n\_estimators=200, max\_features=6, max\_depth=6 \\
LinearSVC* & 44 & 142 & 4 & 24 & 92 & 38 & 76 & C=20, tol=0.0002\\
SVC* & 36 & 37 & 12 & 49 & 75 & 59 & 81 & kernel='rbf', C=25, gamma=0.02\\
LogisticRegression* & 42 & 128 & 6 & 25 & 88 & 39 & 78 & C=60, tol=0.0006\\
KNN* & 45 & 151 & 3 & 23 & 94 & 37 & 74 & n\_neighbors=7, weights='distance'\\
RandomForest* & 45 & 137 & 3 & 25 & 94 & 39 & 78 & n\_estimators=200, max\_features=6, max\_depth=6 \\
 
\hline
\end{tabular}
\caption{Results of the classification algorithms, trained on 4,697 samples. The results here are reported for the 2,314 samples making up the test dataset. We report here the planets correctly identified (True Positives - TP), non-planets incorrectly labeled as planets (False Positives - FP) and the number of planets missed by the algorithm (False Negative - FN). From this we calculate the precision (TP/(TP+FP)), recall (TP/(TP+FN)), and F1 score (F1 = 2*Precision*Recall/(Precision + Recall)) for the planets. The accuracy reflects the performance of the classifier as whole, and shows the total number of correct predictions of any label divided by the total number of samples in the test dataset. The full confusion matrix for all of the classifiers reported here can be found as an appendix in the online journal. The Tuned Parameters column reports the best performing values for the listed parameters, as determined by using the GridSearchCV function, which tests all combinations of a specified grid of parameters. Classifiers marked with an * use a training sample that added training examples using SMOTE sampling, explained more fully in text. By adding SMOTE datapoints, the training set grew to 10,672 samples, while the testing dataset remained the same. The Linear SVC, SVC, and Logistic Regression Classifier that did not use the SMOTE datapoints are run using the keyword class\_weight='balanced', which weights each sample based on the number of entries in the training dataset with that label. The Random Forest classifier without SMOTE resampling uses the class\_weight keyword 'balanced\_subsample', which has the same function but is recomputed for each bootstrapped subsample. While the SVC both with and without the SMOTE training sample performed very well in the 'f1' score and overall accuracy, the false negative rate was higher than desired for practical application. The RFC showed the highest recall, which we want to maximize to find the widest range of planets. We therefore prefer the RFC,  with a trade-off of having more false positives that would be flagged for follow-up. }
\label{table:classification_algorithms}
\end{table*}

\subsection{Random Forest Classifiers} \label{RFC;}
After exploring several different classification techniques, we decided to pursue the RFC in more detail because of its high recall rate (See Table \ref{table:classification_algorithms}). RFC is one of the most widely used machine learning techniques, and is particularly useful in separating data into specific, known classes. Applications to astronomy include tracking the different stages of individual galactic black hole X-ray binaries by classifying subsections of the data \citep{Huppenkothen2017}, classifying the source of X-ray emissions \citep{Lo2014}, and distinguishing between types of periodic variable stars \citep{Masci2014,Rimoldini2012}. 

RFC has many advantages, most notably the ease of implementation, solid performance in a variety of applications, and most importantly for our study, the easily traceable decision processes and feature ranking. It is because of this last advantage that we focus our attention on the RFC over the linear SVC, SVC, linear regression, or KNN methods. RFC is an ensemble method of machine learning, comprised of several individual decision trees. Each decision tree is trained on a random subset of the full training dataset. For each `branch' in the tree, a random subset of the input characteristics known as `features' are selected and a split is made based on a given measure that maximizes correct classifications, with samples falling above and below the split point moving to different branches. Each branch then splits again based on a different random subsample of features. This continues until either all remaining items in the branch are of the same classification or until a specified limit is reached. The output of the RFC is a fractional likelihood that the input object falls into each category, and the highest likelihood is then used as the classification.


On its own, each decision tree does not generalize well to other datasets to make predictions. However, by training many different trees and combining them to use the most popular vote as the prediction, a more successful and generalized predictor is created. There are many ways to tune the RFC to try to optimize the classification for the dataset. For example, the number of trees in the forest, the depth of the tree (the number of splits that can be performed), the number of features available at each split, and the method used to optimize the splits are all characteristics that can be modified. There is no set rule for choosing these parameters, and therefore many different tests are conducted to try to optimize the results. Here, we use \textsc{Scikit-learn's} RandomForestClassifier method to perform the classification and to tune the parameters. We find that the highest-performing forest contains 200 trees, above which the accuracy increased very minimally for the increased computation time. Each tree was capped at a depth of 6 and allowed 6 random features at each split in the tree, which helps to reduce overfitting. 


One advantage of the RFC method is that many different features can be included, and not all features need to be effective predictors. This can be useful for exploratory data analysis as the user can include all of the various data elements without introducing biases from curating the input list. Conversely, it is important to note that the performance of the RFC depends on the quality of the input features. It is possible that the performance of all classifiers could increase if better features are identified.

\begin{figure*}
\centering{\includegraphics[scale=.6]{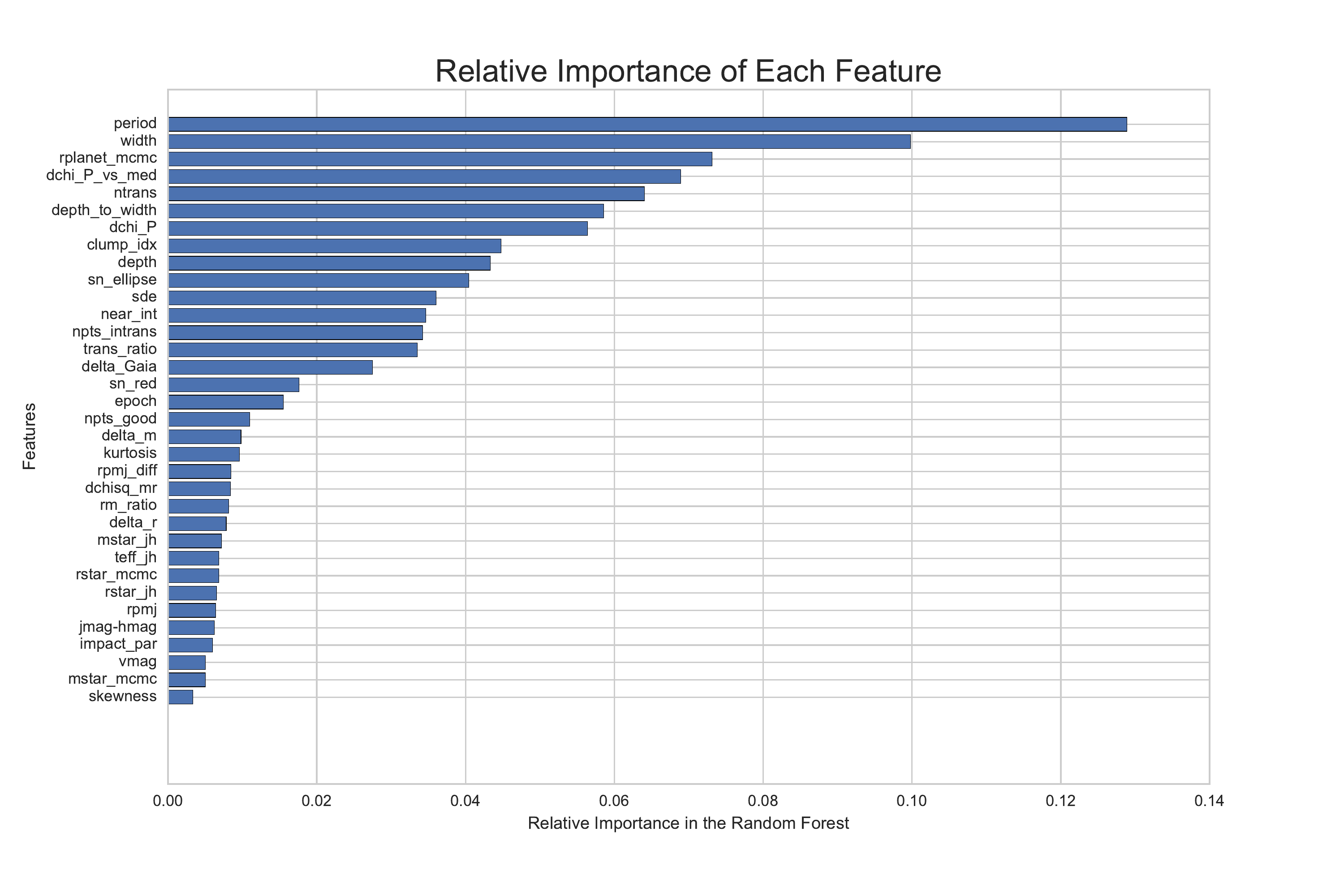}}
\caption{Ranked list of the effectiveness of each of the features in making correct classifications of the training dataset for the Random Forest Classifier. Properties of the transit itself, such as the period and width, are shown to be important discriminators in identifying genuine transits, while stellar properties like magnitude and mass are not effective for classification. } 
\label{fig:RF_feature_importances}
\end{figure*}

As a byproduct of the training process, the RFC can analyze the importance of the various features in making predictions. The results of such an analysis using our training dataset are shown in Fig. \ref{fig:RF_feature_importances}. This can be used to gain insight into the decision making process that the classifier has developed, which can inform further analysis. For example, the period of the planet was the strongest indicator. This can be explained in large part because false planet detections arising from diurnal systematics tend to have orbital periods close to multiples of one sidereal day due to the day/night cycle present in Earth-based observations. This indicator would likely not play as significant a role in a space-based survey unaffected by the day/night cycle. The width or duration of the transit, estimated radius of the planet, the $\Delta\chi^2$ value (a product of the BLS search) of the object at the best-fit period, and the number of transits of the object round out the top 5 features in prediction.

The features that had relatively little impact on the overall prediction related largely to stellar properties, including the magnitude and radius of the star. This shows that there is no strong preference for a certain size star to host a particular type of object in our sample, as the apparent magnitude range to which WASP is most sensitive is dominated by F and G stars \citep{Bentley2009PhD}. The lowest ranked feature is the skewness, showing that the asymmetry of datapoints falling within the best-fit transit is not sufficiently capturing the transit shape information.

\begin{figure*}
\centering{\includegraphics[scale=0.5]{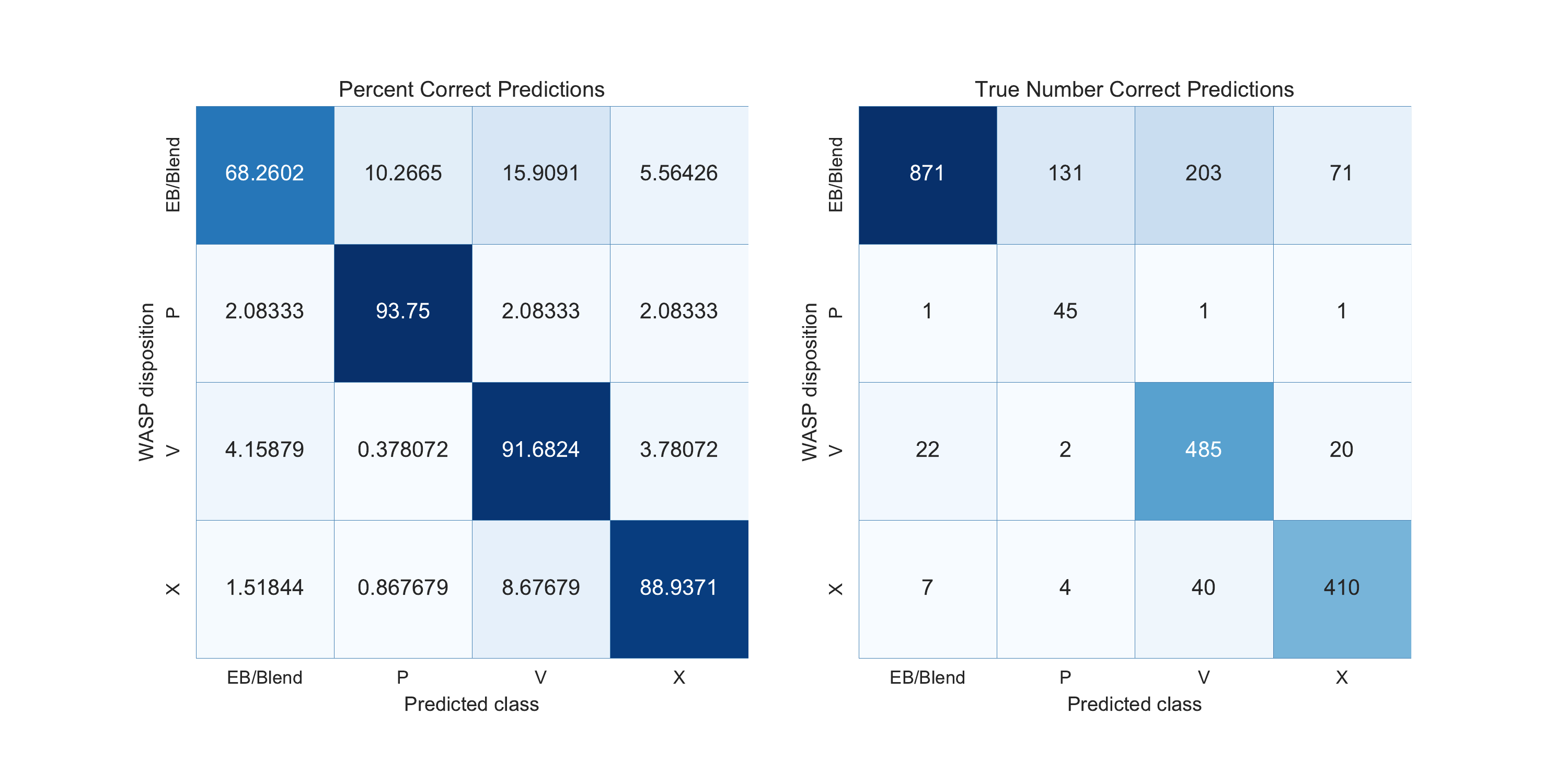}}
\caption{Confusion matrix showing the results of the RFC using the training dataset containing synthetic datapoints generated through SMOTE sampling. The x-axis indicated what the algorithm predicted and the y-axis displays the human labeled class, which we assume to be accurate.  Correct predictions fall on a diagonal line from upper left to lower right. The plot on the left shows the results as a fraction of lightcurves that fall into that bin. However, since the number of samples in each class varies, a more practical depiction is shown in the right plot, which shows the actual number of lightcurves for each category. }
\label{fig:RF_confusion}
\end{figure*}

The results of the RFC, shown in Fig. \ref{fig:RF_confusion}, show that $\sim$93\% of confirmed planets are recovered from the dataset. The ones that are not recovered are most often rejected with the label 'X'. The more concerning statistic is that more than 10\% of EB/Blends were misclassified as planets.  Since there are far more binary systems recorded than there are planets, this quickly turns into a large number of lightcurves incorrectly identified, which translates to many hours of wasted follow-up time.  For our testing dataset, there are 45 correct planet identifications and 137 that are false positives. This means that if all objects flagged as planets are followed up on, we would expect about 25\% of them to be planets. In reality, not all flagged objects are good candidates and can be eliminated by visual inspection. Regardless, the high false positive rate led to the development of Convolutional Neural Networks as a secondary test on the lightcurves, as discussed in \ref{CNN}. 

%
%

\subsection{Neural Networks}
The past decade has shown an explosion of new applications of Neural Networks to tackle a variety of different problems. The particular flavor of neural networks is dependent on the specified task at hand and the types of data available for training. The basic idea of a Neural Network was originally inspired by the neurons in a human brain, although in practice artificial neurons are not directly analogous. Nonetheless, like the human brain, this type of structure is useful for learning complex or abstract information with little guidance from external sources. In the following section, we discuss Convolutional Neural Networks (CNN) and their application to the WASP database. 

\subsubsection{Convolutional Neural Networks} \label{CNN}

\begin{figure}
\centering{\includegraphics[width=\columnwidth]{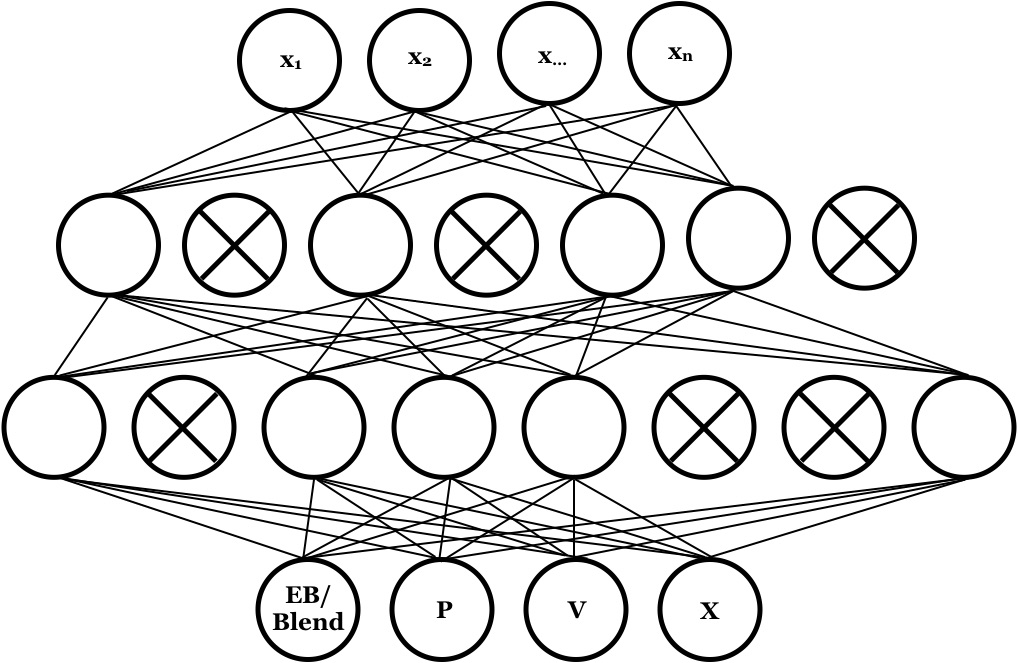}}
\caption{Visual representation of a neural network scheme, where circles represent individual neurons. In this example, the layers progress from top to bottom. The top row represents the input layer, followed by 2 hidden layers, and finally an output layer, where the class prediction is made.  Circles with crosses through them represent dropped neurons, described further in text. } 
\label{fig:neural_network}
\end{figure}

A standard Artificial Neural Network (ANN) has the basic structure of an input layer containing the features of the input data, one or more hidden layers where transformations are made, and the output layer which offers the classification. The output of each layer is transformed by a non-linear activation function. The basic building-blocks of ANNs are known as neurons. A basic schematic of the system architecture is shown in Fig. \ref{fig:neural_network}.

In this work, we use the \textsc{Keras} package \citep{Chollet2015} to implement the neural network, which offers a variety of built-in methods to customize the network. At each layer, the input data is passed through a non-linear activation function. Many activation function choices are available, but here we choose the rectified linear unit, or 'relu' function \citep{Nair2010} for all layers except the output layer, to which we instead apply a sigmoid function. The relu function was chosen both because of its wide use in many applications and because of its high performance during a grid search over our tuning parameters. In addition to the activation function, each neuron is assigned a weight that is applied to the output of each layer, and is modified as the algorithm learns. Using \textsc{Keras}, we tested a random normal, a truncated normal distribution within limits as specified by LeCun \citep{LeCun1998b}, and a uniform distribution within limits specified by He \citep{He2015} initialization of the weights and found the greatest performance with He uniform variance scaling initializer.

The learning itself is managed using the Adamax optimizer \citep{Kingma2014}, a method of optimizing gradient descent. In gradient descent, classification errors in the final output layer are propagated backwards through the network and the weights are adjusted to improve the overall error during the next pass through the network. This is an iterative process, with small adjustments made after each pass through the network until a minimum error (or maximum number of iterations) is reached. The maximum change in the weights allowed by Adamax at each iteration is controlled by the learning rate, which can be tuned to different values for individual datasets. Finally, at each stage, we incorporate neuron dropouts, where a fraction of the neuron inputs are set to 0 in order to help prevent overfitting. The total dropout percent is determined experimentally, and here we find 40\% to be effective.

While the numerical data used in the RF could be passed to the neural network directly, this data only provides insight into the statistical distribution of the classes, producing results similar to that of the RFC. In practice, it can be the case that many of the features fall within the right range to be labeled with a certain class, but a quick visual inspection of the lightcurve can easily rule the classification out. Instead it would be desirable to create an algorithm that can use the shape of the lightcurve itself to make classification decisions. Using the lightcurve data would more closely mimic the human process of eyeballing.

\begin{figure}
\centering{\includegraphics[width=\columnwidth]{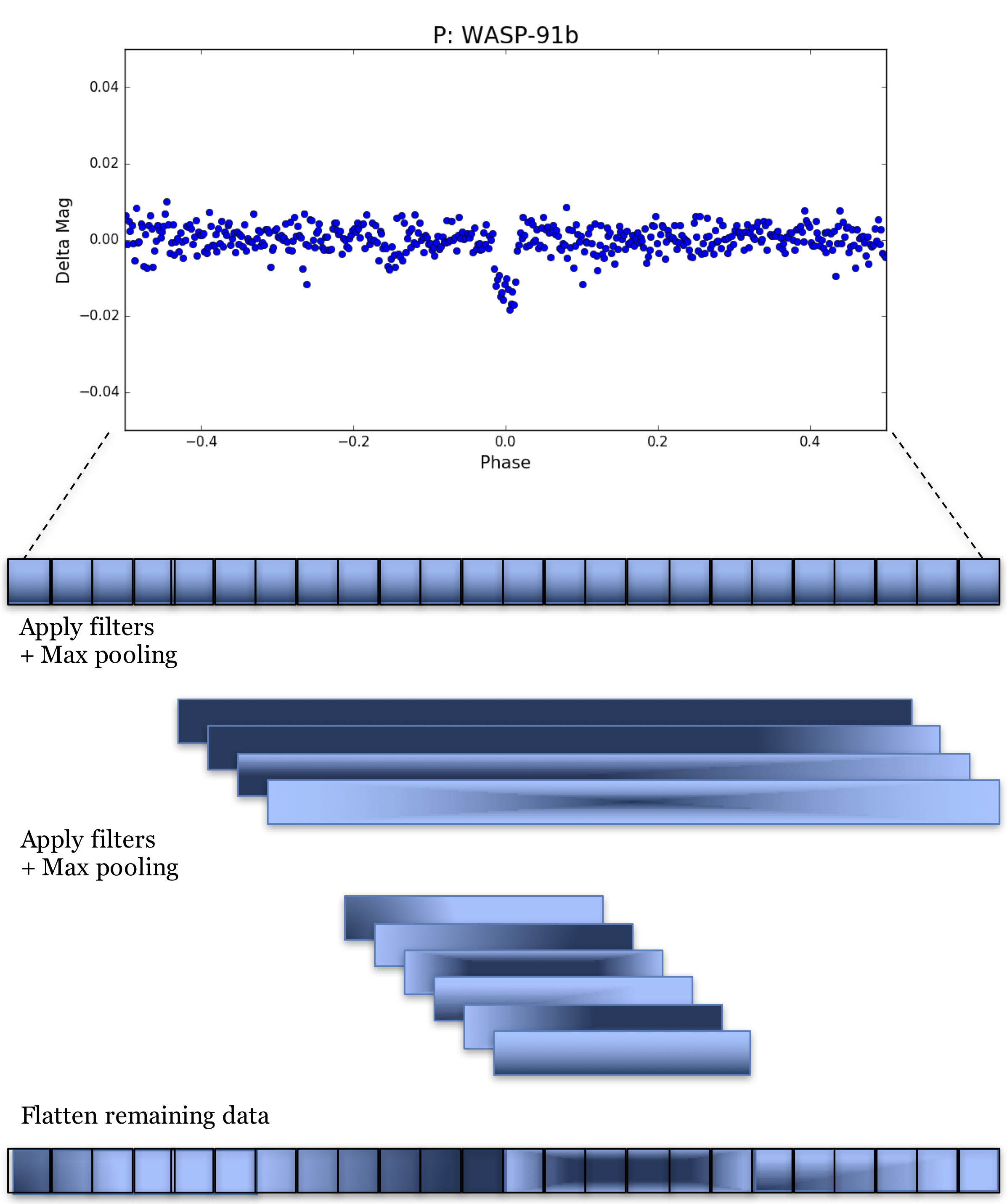}}
\caption{Simplified visual representation of the convolution steps for a full lightcurve, folded on the best fit period. The shading of the boxes represents the different filters that are applied to the data in each step. Also at each convolution step, the data length is reduced using the Max Pooling method, in which only the maximum value of every $n$ datapoints is kept. In the final step, all of the pooled and filtered data are stacked one after the next. This new one-dimensional dataset is then passed into the fully connected layers of the neural network. The final network architecture is as follows: the input data is comprised of one dimensional magnitude data binned to length 500. The first convolution step has 8 filters with a kernel size of 4. The data is then pooled by 2. Each lightcurve at this point has 8 layers of length 250. The second convolution step has 10 filters with a kernel size of 8, followed again by another pooling layer with a pool size of 2 leading to a datasize of 10 layers of length 125. The information from all of the filters is then flattened into one layer of length 1,250 for each lightcurve, and is input to the first densely connected layer. There are 2 dense hidden layers of size 512 and 1024. Each stage of the convolution and fully connected layers has a 40\% dropout rate to prevent overfitting. The final output layer has 4 neurons, one for each classification category. The CNN with both the full and local lightcurve has the same general format, but with the input data being a stack of two 500-length lightcurves. }
\label{fig:convolution}
\end{figure}

In order to directly use the lightcurve data, we developed a Convolutional Neural Network \citep[CNN;][]{LeCun1990, Krizhevsky2012}. In this case, the input data we use is the actual magnitude measurements of WASP folded on the best-fit period determined by the BLS. In the CNN that we adopt, the magnitude data first undergoes a series of convolution steps, in which various filters are applied to the data to enhance defining characteristics and detect abstracted features. The filters themselves are optimized iteratively to find those that best enhance the differences between classes, similar to the way the weights between the neural layers are updated. The convolution process is represented in Fig.\ref{fig:convolution}.

Each WASP field contains a different number of sample points, yet input to the CNN for all of the samples need to be of the same length. To standardize the data length, the folded lightcurves are divided into 500 bins in transit-phase space. The number of bins was determined experimentally, with 500 being the best trade off between providing detailed lightcurves without having a significant number of bins missing data.

Once the data are binned, a set of one-dimensional convolution filters are passed over each lightcurve. This has the effect of making our dataset larger by a factor of the number of filters, so to reduce the data size and help prevent overfitting we apply a MaxPooling layer where we only keep the maximum value of every n data points. Finally a dropout layer is added, in which a random specified fraction of the points are ignored to prevent overfitting the training data. We then repeat this entire process to add more complex and abstract filters. Finally, all of the remaining data is flattened so that each lightcurve, now comprised of several filtered and pooled representations of the original lightcurve, are added together to make a one dimensional array for each star, which is then passed into the fully connected layers of the neural network. All of the layers of the network are optimized to provide the best fit to the data in the final classification layer. 

The training set is highly unbalanced with relatively few planets as compared to eclipsing binaries and blends, variables, or lightcurves with no signal, which limits the performance of the CNN. To compensate, we extended the sample of 'P' training examples by creating artificial transit lightcurves.  

The lightcurve injection was done by adding synthetic transit signals to existing WASP light curves that showed no transit signal or other variability. This ensured realistic sampling and typical patterns of correlated and uncorrelated noise. We began with a sample of lightcurves of objects classified as 'X' in the Hunter catalog, meaning they have been rejected and contain no  detectable planet signal. We started with all X stars and measured the RMS against the V-magnitude. Those objects that fell more than $1\sigma$ below the best fit to the data were selected, as they show the least amount of variation. This left a total of 848 light curves.

The planetary signal added to the WASP data was created using the \textsc{batman} package for Python \citep{Kreidberg2015}.  The stellar mass, radius, and effective temperature are set using the known values for the star itself. The planetary properties were generated randomly with the following distributions.

The period was randomly selected to be a value uniformly located in log space between 0.5 and 12 days, as this is the range in which the WASP pipeline typically uses the BLS algorithm to look for planets. The mass of the planet follows the same lognormal distribution used in \cite{Cameron2018}, with a mean of 0.046 and a sigma value of .315. The semi-major axis can be found for the period using 

\begin{equation}
a = (\frac{p^2G(M_1+M_2)}{4\pi^2})^\frac{1}{3}
\end{equation}

The radius of the planet is dependent on both the mass of the planet and the equilibrium temperature. We use a cubic polynomial in log mass and a linear term in log effective temperature to approximate the planetary radius, using coefficients derived from a fit to the sample of hot Jupiters studied by \cite{Cameron2018}:

\begin{eqnarray}
\log \left(\frac{R_p}{R_{\rm Jup}}\right)&=&
c_0
+c_1\log\left(\frac{M_p}{0.94 M_{\rm Jup}}\right)
+c_2\log\left(\frac{M_p}{0.94 M_{\rm Jup}}\right)^2\nonumber\\
&&+c_3\log\left(\frac{M_p}{0.94 M_{\rm Jup}}\right)^3
+c_4\log\left(\frac{T_{\rm eql}}{1471\ {\rm K}}\right)^4\label{eq:radfit},
\end{eqnarray}
where $c_0=0.1195$, $C_1=-0.0577$, $c_2=-0.1954$, $c_3=0.1188$, $c_4=0.5223$, and $T_{eql} = T_{eff}\left(\frac{R_S}{2a}\right)^\frac{1}{2}$ 

As we are looking only for close-in planets, we make the simplification that all eccentricities are 0. Finally, the inclination was calculated by first randomly picking an impact parameter, b, between 0 and 1. The inclination was then calculated by

\begin{equation}
i = \cos^{-1}\left(\frac{bR_S}{a}\right)
\end{equation}

The lightcurves were generated and added to one of the selected WASP lightcurves folded on the assigned period. While some of these new planets were too small to be visible, and others were much larger than would be expected, we chose to include them all in order to push the boundaries of the parameter space that the CNN is sensitive to so as not to exclude potentially interesting, though unusual, objects.

The lightcurves of the artificial planets, as well as real-data examples of P's, EB/Blend's, V's, and X's are phase folded on the best-fit period (or assigned period, in the case of the artificial planets) and binned by equal phase increments. Including the artificial planets, there are 4,627 objects in the training set and 2,280 in the testing dataset.

The CNN parameters were set by using a grid search over the tunable parameters. The final CNN was comprised of two convolutional layers with 8 and 10 filters respectively. The pooling stages each had a size of two, and 40\% of the neurons were dropped at each set. The flattened data were passed to a neural network with two hidden layers of sizes 512 and 1048 and with a 'relu' function. Both of these layers also had a 40\% dropout applied. The learning rate for the Adamax optimizer was 0.001. The most effective batch size was 20 with 225 total epochs. As with the RFC, the output of the CNN is a likelihood that the lightcurve falls into each of the categories, with the highest likelihood being the prediction. 

Using only the binned lightcurve as input, the CNN achieves an overall accuracy (correct predictions divided by total lightcurves analyzed) of around 82\% when applied to the test dataset. While the fraction of correctly identified planets is lower than the RF (88\% as opposed to 94\%), the CNN performs much better in classifying eclipsing binaries and blends in terms of the percent of false positives with only $\sim$5\% of EB/Blends being labeled as planets, as opposed to 10\% for the RF. The CNN therefore has an overall better performance for follow-up efficiency.

\begin{figure}
\centering
   \includegraphics[width=\columnwidth]{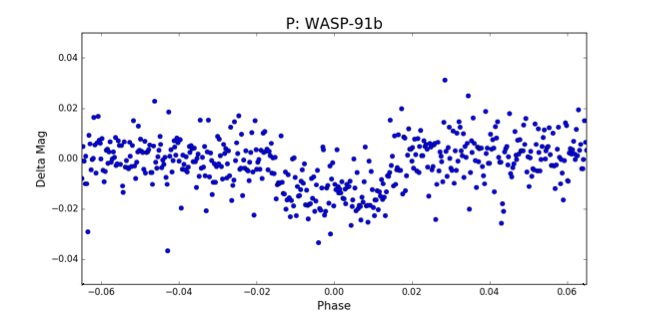}
   \caption{A local view of WASP-91b, the same planet as shown in the CNN example (Fig. \ref{fig:convolution}). This local view was used as an additional layer in the second CNN. The local version clearly shows more detail on the transit shape, and specifically the flat-bottomed transit. EB/Blends tend to have a 'V' shape with steeper sides, which is more apparent in this close-up view. }
   \label{fig:WASP_91_local} 
\end{figure}

\begin{figure}
\centering
   \includegraphics[width=1\linewidth]{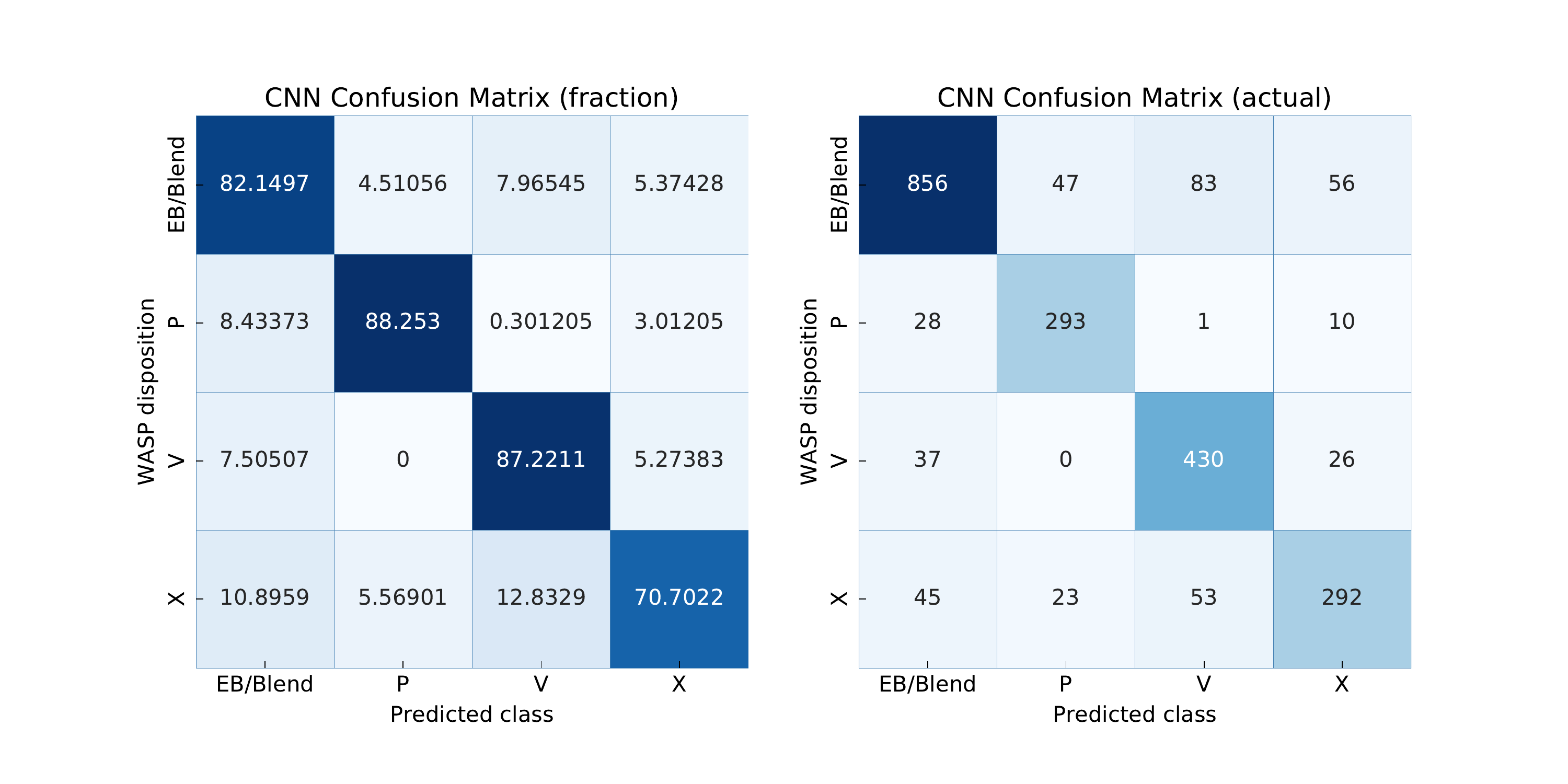}

   \includegraphics[width=1\linewidth]{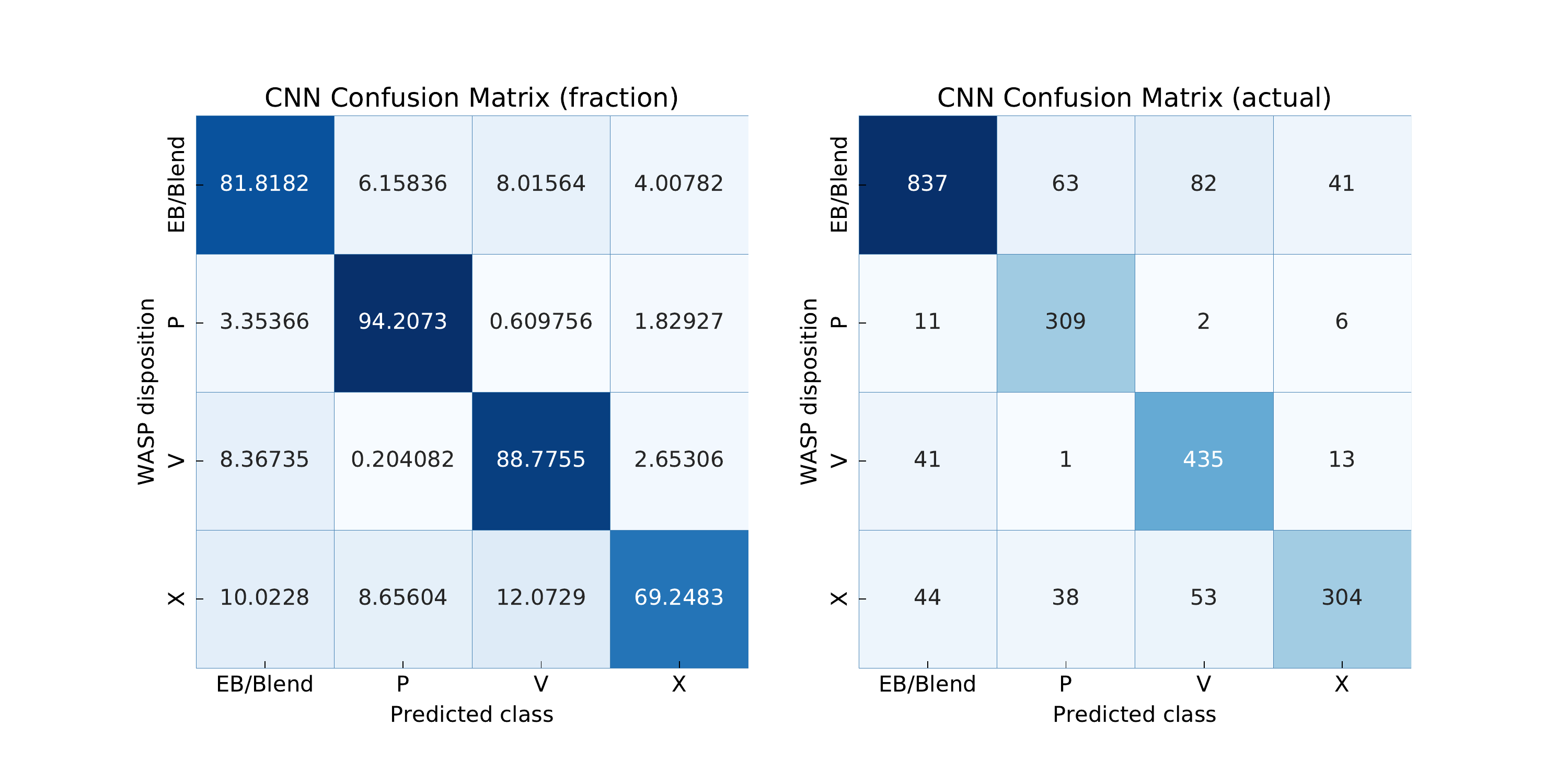}
\caption{Confusion matrix showing the results of the CNN using only lightcurves folded on the best-fit period (top) and with the addition of the local transit information (bottom) as input. The axes are interpreted the same as in Fig. \ref{fig:RF_confusion}. The plot on the left shows the results as a fraction of lightcurves that fall into that bin. The right plot shows the actual number of lightcurves for each category. Note that in this example, we artificially injected additional planets into WASP data to increase the sample, so the numbers reported are for a combination of the real and artificial planets.}
\label{fig:CNN_confusion_matrices}
\end{figure}

In order to further increase the performance of the CNN, we train a second CNN algorithm to include the local transit information, using an approach similar to that of \cite{Shallue2018}. The local information is comprised of the data centered on the transit and only containing the data 1.5 transit durations before and after the transit event, standardizing the transit width across events. An example of a local lightcurve is shown in Fig. \ref{fig:WASP_91_local}. The effect of this is to provide greater detail and emphasis on the shape of the transit event itself in order to understand the subtle shape difference between a typical planet and an eclipsing binary system. The full lightcurve and the local view are stacked and passed together into the CNN. In this case, the overall accuracy (83\%) remained roughly the same, but the total percentage of planets found increased to (94\%). The trade-off is a slight increase of the number of EB/Blends being labeled as planets. A full comparison of both methods can be seen in Fig. \ref{fig:CNN_confusion_matrices}.

It is important to note that the way missing data was handled for both the full lightcurve and the local lightcurve made a large difference in the final performance. When binning the data, the full dataset was evenly split in 500 equal phase steps and all of the datapoints within those phase steps were averaged. In some cases, for example when the lightcurve was folded over an integer day, there were gaps in the phase ranges in which no data were present. Since the CNN can not handle missing data in the input string, a value needs to be inserted. We tried inserting either a nonsense value, in this case 0.1 which is far above any real datapoint, or repeating the last good value. In some cases there were several phase steps in a row that were missing data, causing a small section of the lightcurve to be flat. 

After trying both options, we found that by far the best performance was obtained when inserting the nonsense value into the full lightcurve and repeating the last good value into the local lightcurve. This makes sense, as the full lightcurve gives a broader view of the star's lightcurve and is likely to have regular gaps in the data when it is folded on a bad period. The algorithm was able to identify that pattern and reject it. The local data, however, have fewer total datapoints because they cover a smaller total range of phases, and therefore are more likely to randomly have missing data. The algorithm is no longer able to distinguish lightcurves missing data because of an intrinsic problem with the data fold and those missing data simply because they lack enough observations during the transit, confusing the results. 

\section{Analysis and Results} \label{Analysis}

When looking at the results of the RFC and CNNs, the percentage of correct predictions across methods is consistent, with $\sim90\%$ of planets being correctly identified. However, when looking at the original lightcurves for both true positives and false negatives, clear patterns in the different machine learning methods begin to emerge.

\begin{figure*} 
\centering{\includegraphics[scale=.45]{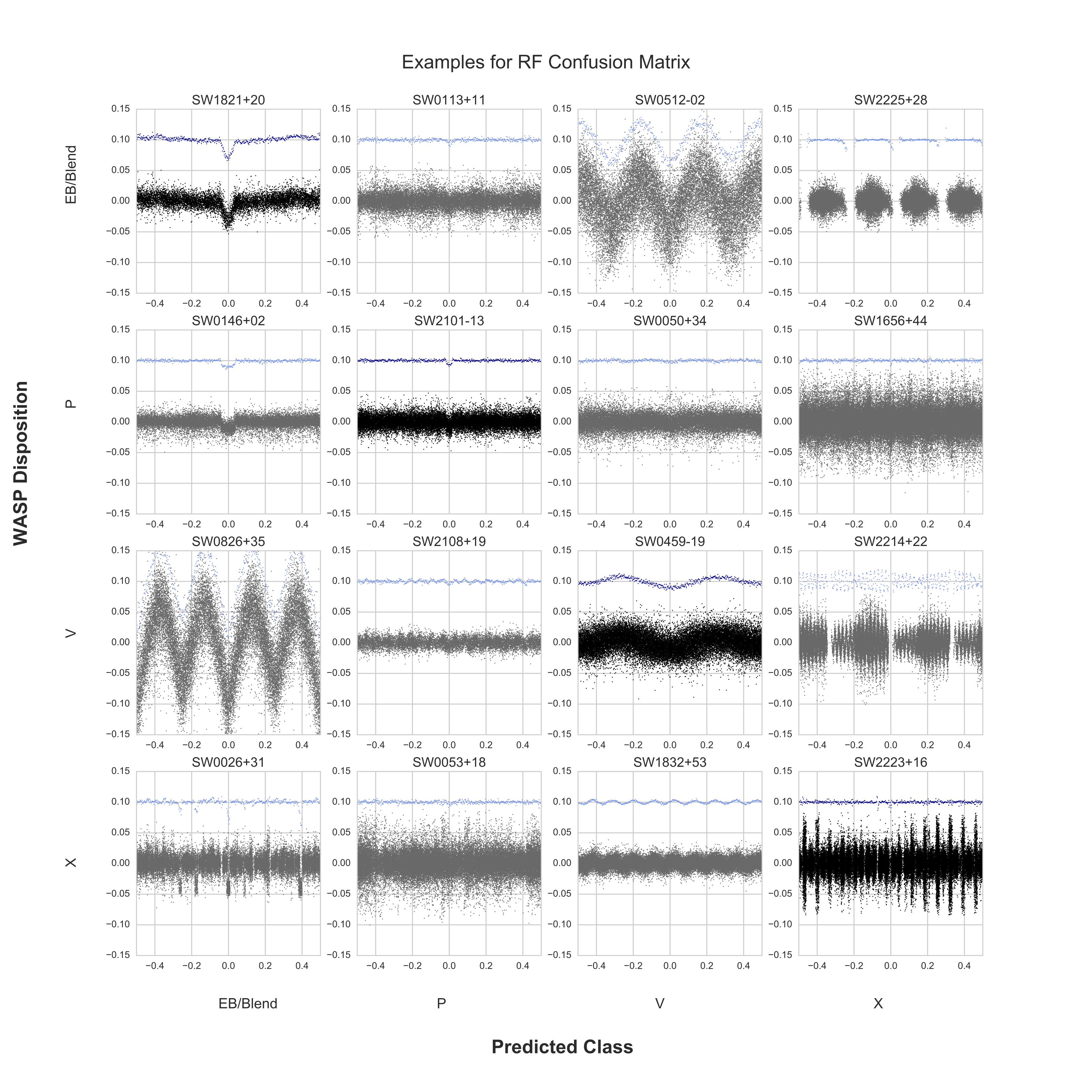}}
\caption{Confusion matrix of RFC results showing examples of lightcurves selected from samples that fall into each category, chosen to represent typical failure modes. Lightcurves along the diagonal, shown in black, were correctly classified by the RFC. Off diagonal boxes, shown in gray, were incorrectly identified, with the true classification shown on the vertical axis and the predicted classification shown on the horizontal axis. Looking at the samples in the off-diagonal boxes provide insight into how the RFC makes its decisions and what the common failure modes are. SW1832+53 was labeled as an X in the archive, but the RF predicted it was a variable object. This classification was made early in WASPs' history, and a clearer picture of the lightcurve has since been established. While an X classification means that there is no planetary signal, a better classification for the object would be to label it as a variable lightcurve, which is what the random forest does. SW0826+35 is another interesting object. It was labeled as a variable in the archive, but the alternating depths of dips indicates it may actually be an eclipsing binary, consistent with the machine classification. The final object of special note is SW0146+02, verified as WASP-76b. This planet's transit is particularly deep and confused the RFC into mislabeling it as an EB/Blend showing that this algorithm is sensitive to the depth of transit despite the classical 'U' shape of the event.  } 
\label{fig:RF_cmtx_examples}
\end{figure*}

\begin{figure*}
\centering{\includegraphics[scale=.45]{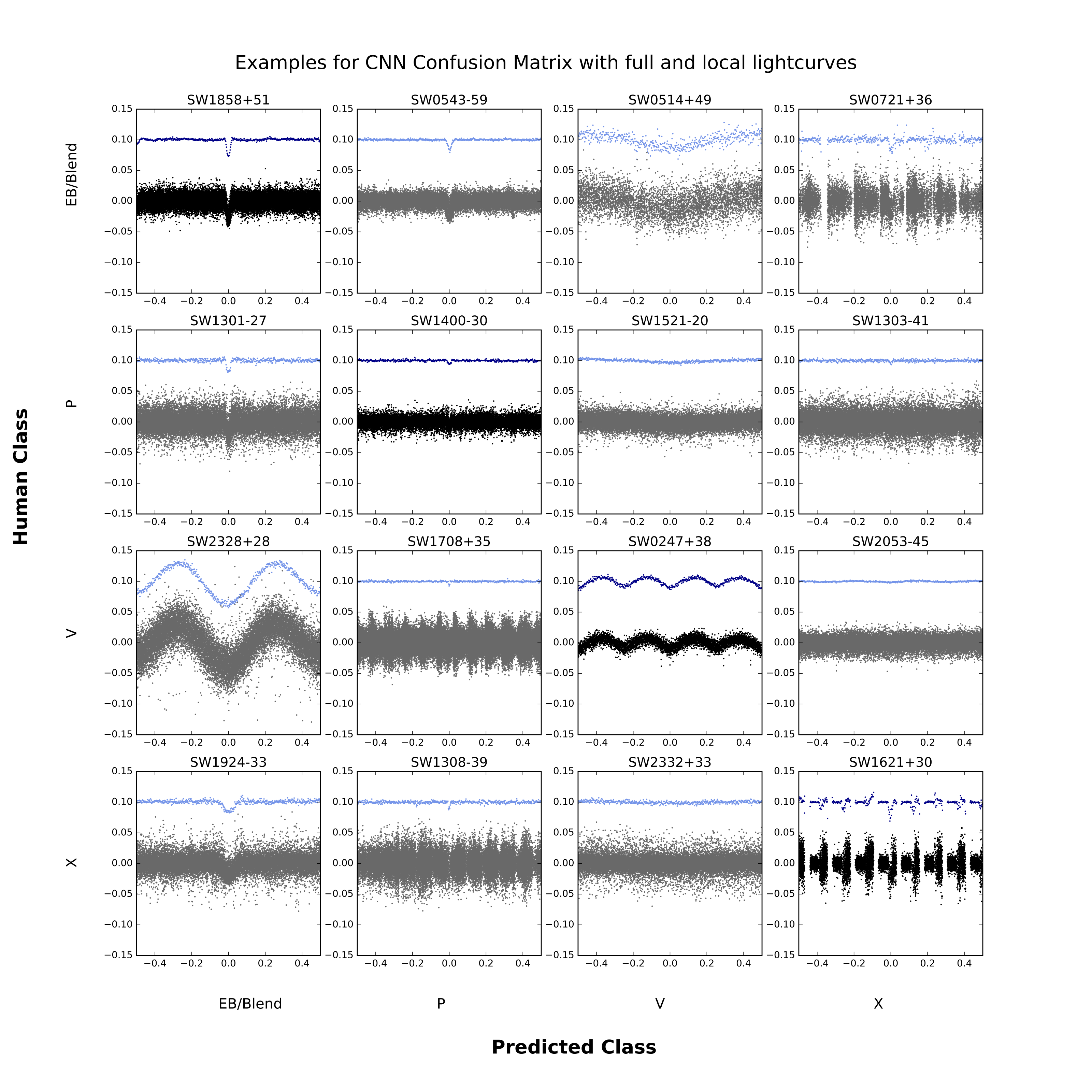}}
\caption{Same as for Fig. \ref{fig:RF_cmtx_examples}, but for the CNN using the local and full binned lightcurve. As in the RFC, the overlap in categories in the human classifications is evident in the CNN results. For example, SW1924-33 was labeled by a WASP team member as an X because it does not contain a planetary transit, but it clearly does show a transit event and therefore could be instead classified as an eclipsing binary. While this is considered a wrong classification in the algorithm evaluation, in practice it is an acceptable output. SW1308-39 is an example where near-integer day (in this case near 11 days) effects can look like transit signals when the data is phase binned. The CNN did miss several true planets, such as SW1303-41 (WASP-174b; \citealt{Temple2018}), as the dip is very small with a messy lightcurve. SW1521-20 is an example of a planet found in a different survey (EPIC 249622103; \citealt{David2018}) with the signal not being visible in the WASP data.} 
\label{fig:CNN_2d_cmtx_examples}
\end{figure*}

The RFC uses features that are derived from the fitted light-curve parameters and external catalog information, but the lightcurves themselves are not included. This logically leads to candidates that have typical characteristics of known exoplanets to emerge. However, because the WASP data can be very noisy and have large data gaps, there are many occasions where the derived 'best fit' planet features fall into the known distribution, but upon inspecting the original data it is clear that there is no periodic transiting signal. Examples of true positive and false negative classifications for the RFC are shown in Fig. \ref{fig:RF_cmtx_examples}. The main contributors of false positives for the RF is the blended star (rather than the eclipsing binary) component of the EB/Blends. In many cases, the blended stars look very similar to planets by their numerical descriptors, and in particular the depth of the transit and the distribution of transit durations look very plausible. Without looking at the nearby stars, these objects are very hard to distinguish. 

The CNN has a fundamentally different method of identifying transits. As described in section \ref{CNN}, the CNN is not provided with derived data, but rather has direct access to the magnitude data folded on the best-fit period. In this case, the algorithm is simply trying to pattern-match to find the correct shape for a transit. Looking at the true positives and false negatives (examples shown in Fig. \ref{fig:CNN_2d_cmtx_examples}) for this subsection of data shows a different failure mechanism for wrongly-identified planets. In many cases, light curves will look like planets, but when other information, such as the depth of the transit, is known it becomes clear that the object is more likely an eclipsing binary or other false positive. In addition, fainter objects tend to have much noisier data and more sporadic signals, which can sometimes look like a transit signal when the data is binned down to 500 data points. Finally, the drift of stars across the CCD during each night can lead to systematic disturbances that are consistent at the beginning or end of each night in some (but not all) target stars. Since this effect is specific to each star, it is not always corrected by decorrelation. This can lead to the lightcurve having a clear drop in magnitude at regular intervals, and the gaps in the data can appear transit-like to the CNN. Interestingly, this last problem is far more prevalent when fewer neurons in the ANN are used. Increasing the neurons to 512 and 1024 in our final configuration nearly eliminates the problem, although a few cases do still get through.

\section{Discussion} \label{Discussion}
Each of the machine learning methods above performs best on a specific subset of planets. Like a human looking at a list of transit properties, the RFC and SVC are better suited to finding planets that have strong signals and have properties similar to those of the other known WASP planets. The CNN using the magnitude data folded on the best fit period functions similar to a human eyeballing a lightcurve and making decisions. By combining the predictions of these methods, we get a more robust list of planetary candidates. The importance of the combination of machine learning algorithms has also been noted by others \citep{Morii2016, Disanto2016}, and will be an important framework for upcoming large-data surveys.  

Currently, radial velocity follow up of WASP targets takes place primarily using the CORALIE spectrograph at the La Silla Observatory in Chile \citep{Queloz2000} for southern targets and the Spectrographe pour l’Observation des Phénomènes des Intérieurs stellaires et des Exoplanètes \citep[SOPHIE;][]{Perruchot2011} at the Haute-Provence Observatory located in France for targets in the north.

Since thorough records have been kept of the WASP follow-up program with CORALIE, 1234 candidates have been observed and dispositioned. Of those, 150 (12\%) have been classified as planets (2 of which are the brown dwarfs WASP-30 and WASP-128), 713 (58\%) are binaries or blends, 225 (18\%) were low mass eclipsing binaries, and the remaining 146 (12\%) were rejected for other reasons, including 60 because the stars turned out to be inflated giants. The SOPHIE follow-up effort has a similar success rate to date. Of the 568 total candidates dispositioned, 53 (9\%) are planets, 323 (57\%) are blends or binaries, 116 (20\%) were low-mass eclipsing binaries, 72 (13\%) were rejected for other reasons including being a giant star, and 4 (1\%) were variable stars.

As a comparison, for our RFC, 182 objects were classified as planets, with 45 true positives and 137 false positives, indicating a success rate of 25\%. The SVC is more conservative, finding fewer total planets but rejecting more false positives, and has a 49\% estimated follow-up accuracy (True positives divided by true positives and false positives). The CNN with the full lightcurve showed even better results, with 81\% estimated follow-up accuracy, and when the local lightcurve data was added 75\% of the objects flagged as planets were true positives. It may seem like the CNN would be the best option to use alone, but it did miss several planets that the RFC was able to recover and occasionally let in false signals that were caught by the RFC or SVC, highlighting the importance of combined methods. 

We note that the true follow-up rate for any of these methods would be lower in practice. Eclipsing binaries with low-mass stellar companions, which closely resemble planets in their lightcurves and derived features, were removed from the training dataset. When we applied the RFC to 399 low-mass eclipsing binary systems, 92 (23\%) were classified as planets. The CNN with only the full lightcurve also returned 64 as planets, partially overlapping with the RFC predictions. Adding the local lightcurve information made the CNN more likely to identify EBLMs as planets with 95 (24\%) being labeled as planets. The SVC is the most shrewd, with only 24 stars labeled as planets. For many of these objects, the transit signals look identical to those of planets and are only discovered with follow-up information. Even by combining results from different machine learning methods, we expect to have many of these type of objects flagged for further observation, reducing the overall performance of the algorithms.

There are several caveats to our study. One note of caution is the underlying training dataset. The training data was obtained by combining the entries of a number of WASP team members over the course of many years. This leads to two main problems. First, different team members may label the same lightcurve differently based on their interpretation. Blends and binaries for example can be used differently by different users. We attempted to control for this by manually inspecting the blends and binaries and updating flags to maintain consistency across all fields. The 'X' category is also notably inconsistent, with objects that were rejected as planets for many different reasons, including blends and binaries, being given the same label. 

The second issue comes from the fact that the classification began before all current data were available. After the first few WASP observing seasons, classifications were made based on the limited data available. When other data were added in the following seasons, the shape of the lightcurve may have changed and more (or less) transit-like shapes became obvious. However, since the candidate was already rejected, it was never re-visited and updated. Several examples like this were found by looking through the incorrect classifications, such as those in Figures \ref{fig:RF_cmtx_examples} and \ref{fig:CNN_2d_cmtx_examples}, and remained uncorrected in our training data. Regardless of these problems, the algorithms were robust and were able to make reasonable predictions even with small variations in the training. 

Finally, we rely on the BLS algorithm to provide an accurate best-fit period. This is especially important for the CNN, which only has the lightcurve folded on that period as input. The CNN is therefore not equipped to handle possible incorrect periods due to aliases or harmonics. It would be possible to augment the code to also include other information for the CNN, such as the data folded on half of the period and twice the period, either in a stack or as a separate entry, to try to identify planets in the data that were found at the wrong period. This possibility will be included in future work, but is beyond the exploratory scope of this paper.

The classifications shown here have a lower accuracy than those reported in the Kepler studies, which range from around 87\% up to almost 98\%. This is to be expected, as WASP data is unevenly sampled and has much larger magnitude uncertainties, making definitive identifications impossible with WASP data alone. WASP's large photometric aperture (48 arcseconds) also makes convincing blends more common. Nevertheless, the machine learning algorithms were able to correctly identify $\sim$90\% of planets in the testing dataset and operate much faster than human observers (less than 1 minute to train the RFC and around 20 minutes to train the CNNs, and less than a minute to apply to new datasets on a MacBook Pro with 3.1 GHz Intel Core i5) and produce more consistent results. The advantage of this is that as new data are added after observing seasons, it is not necessary to look at each lightcurve again. Rather, the entire dataset can be quickly re-run through the algorithms to obtain new observing targets. 

In practice, the machine learning results will be used in combination with expert opinion in order to select the most scientifically compelling targets for follow-up. For example, the area surrounding the star might be crowded with other stars making follow up observations difficult. In several cases, a lightcurve looks promising, but another star within WASP's pixel resolution has already been labeled as a Blend (often through follow-up) and the label did not propagate to the surrounding lightcurves. These are easy to identify manually, but that information is not included for the machine learning algorithms. Therefore it is still essential that targets selected with machine learning are curated by a human user for practical observation. Another factor only taken partially into account by the RFC is the recent improvement in the knowledge of stellar parallaxes, and hence radius estimates, made possible by the first and second data releases of the \textit{Gaia} mission \citep[]{Gaia2016, Gaia2018}. Knowledge of the stellar radius, and therefore the radius of the transiting object, allows clean dwarf/giant discrimination, and eliminates an entire subclass of blended eclipsing binaries at a single stroke.

\section{Conclusions} \label{Conclusions}

While the WASP data alone is not of sufficient quality to definitively identify planets from the data, it has proven to be very effective in producing new candidates for future follow-up and eventual planet status. The large size of the WASP archive makes it undesirable for human observers to manually look at each one to determine whether it is a good candidate for further study. The machine-learning framework we have created provides a tool for the observer wanting to re-examine the full set of data holdings in any WASP field, enabling fast re-classification of all targets showing transit-like behavior and identification of new targets of interest. This list is not intended to be used as a final list for observing, but rather as a tool for the observers to reduce the total number of lightcurves requiring analysis. An additional advantage of this approach is that the algorithms can be quickly re-trained as new information, such as new known classifications from completed follow-up observations, become available. 

Using multiple machine learning models is an effective framework that can be modified and applied to a variety of different large-scale surveys in order to reduce the total time spent in the target identification and ranking stage of exoplanet discovery. Combining the results from additional machine learning methods could further improve the predictions.  

With the launch of the Transiting Exoplanet Survey Satellite (TESS; \citealt{Ricker2014}) and upcoming launch of the PLAnetary Transits and Oscillations of stars (PLATO; \citealt{Rauer2014}), automatic data processing is becoming even more essential. The TESS mission will focus on over 200,000 stars with high cadence ($\sim2$ min) and several million targets with a longer cadence ($\sim30$ min). The PLATO mission will study an additional million targets. The number of lightcurves in these datasets is clearly beyond manual classification, so machine learning techniques will be essential to their success. The application and performance analysis of machine learning on current sky surveys such as WASP are integral to the successful understanding and implementation on future large surveys.

\section*{Acknowledgements}

NS acknowledges the support of NPRP grant \#X-019-1-006
from the Qatar National Research Fund (a member of Qatar Foundation). ACC acknowledges support from STFC consolidated grant ST/R000824/1 and UK Space Agency grant ST/R003203/1. DA, FF, DP, RW and PW acknowledge support from STFC through consolidated grants ST/L000733/1 and ST/P000495/1. DJAB acknowledges support from the UK Space Agency. FF acknowledges support from  PLATO ASI-INAF contract n.2015-019-R0. L.M. acknowledges support from the Italian Minister of Instruction,University and Research (MIUR) through FFABR 2017 fund. L.M. acknowledges support from the University of Rome Tor Vergata through the ``Mission: Sustainability 2016'' fund. DJA gratefully acknowledges support from the STFC via an Ernest Rutherford Fellowship (ST/R00384X/1).  SCCB  acknowledges support  by FEDER - Fundo Europeu de Desenvolvimento Regional funds through the COMPETE 2020 - Programa Operacional Competitividade e Internacionalização (POCI), and by Portuguese funds through FCT - Fundação para a Ciência e a Tecnologia in the framework of the projects POCI-01-0145-FEDER-028953 and POCI-01-0145-FEDER-032113. SCCB also acknowledges the support from FCT and FEDER through COMPETE2020 to grants UID/FIS/04434/2013 \& POCI-01-0145-FEDER-007672, PTDC/FIS-AST/1526/2014 \& POCI-01-0145-FEDER-016886 and PTDC/FIS-AST/7073/2014 \& POCI-01-0145-FEDER-016880 and through Investigador FCT contract IF/01312/2014/CP1215/CT0004.

We acknowledge Prof Coel Hellier for providing detailed information  on the methodology for southern-hemisphere candidate disposition in the later years of the WASP project.



\bibliographystyle{mnras}
\bibliography{exoplanet_papers.bib}

\bsp	

\label{lastpage}
\end{document}